\documentclass [11pt]{article}

\usepackage{jheppub}

\title{Scalar-Fermion Analytic Bootstrap in 4D}

\author[a,b]{Emtinan Elkhidir}
\author[a,c]{and Denis Karateev}

\affiliation[a]{SISSA,Via Bonomea 265, I-34136 Trieste, Italy}
\affiliation[b]{Department of Physics and Astronomy, Uppsala University, SE-751 08 Uppsala, Sweden}
\affiliation[c]{Institute of Physics, EPFL, CH-1015 Lausanne, Switzerland}

\emailAdd{emtinan.osman@physics.uu.se}
\emailAdd{d.s.karateev@gmail.com}

\abstract{
In this work we discuss an analytic bootstrap approach~\cite{Fitzpatrick:2012yx,Komargodski:2012ek} in the context of spinning 4D conformal blocks~\cite{Echeverri:2015rwa,Echeverri:2016dun}.
As an example we study the simplest spinning case, the scalar-fermion correlator $\langle\phi\,\psi\,\phi\,\overline\psi\rangle$.
We find that to every pair of primary scalar $\phi$ and fermion $\psi$ correspond two infinite towers of fermionic large spin primary operators.
We compute their twists and products of OPE coefficients using both s-t and u-t bootstrap equations to the leading and sub-leading orders.
We find that the leading order is represented by the scalar-fermion generalized free theory and the sub-leading order is governed by the minimal twist bosonic (light scalars, currents and the energy-momentum tensor) and fermionic (light fermions and the suppersymmetric current) operators present in the spectrum.
}

\usepackage{graphicx}
\usepackage{color}
\definecolor{red}{rgb}{1,0,0}
\usepackage{cancel}
\usepackage{xcolor}
\usepackage{multirow}
\usepackage{simplewick}
\usepackage[utf8]{inputenc}

\definecolor{red}{rgb}{1,0,0}

\newcommand{\fref}[1]{{\color{gray} \texttt{[#1]}}}

\newcommand{\OO}{\mathcal{O}}
\newcommand{\KK}{\mathbb{K}}
\newcommand{\II}{\mathbb{I}}
\newcommand{\JJ}{\mathbb{J}}

\newcommand{\point}{{\bf p}}

\def\bel{\begin{equation}\begin{aligned}}
\def\eel{\end{aligned}\end{equation}}
\def\be{\begin{equation}} \def\ee{\end{equation}}
\def\zb{\overline{z}}\def\omegab{\overline{\omega}} 
\def\opsi{\overline{\psi}}
\def\oPsi{\overline{\Psi}}

\newcommand{\gammafn}[1]{\Gamma\left( #1 \right)}

\newcommand{\Brk}[1]{ \left( #1\right)}

\newcommand{\ol}{\overline}

\begin{document}
\maketitle
\flushbottom

\section{Introduction}
The conformal bootstrap program~\cite{Mack:1969rr,Polyakov:1970xd,Ferrara:1973yt,Polyakov:1974gs,Mack:1975jr}\footnote{See~\cite{Rychkov:2016iqz,Simmons-Duffin:2016gjk,Penedones:2016voo} for modern introduction.} has developed immensely in recent years since its revival~\cite{Rattazzi:2008pe}.
Many theoretical results maximally exploiting conformal symmetry unveiled rigid kinematic (model independent) structure of correlators in conformal field theories (CFTs)~\cite{Osborn:1993cr,Dolan:2000ut,Dolan:2003hv,Dolan:2011dv,Weinberg:2010fx,Weinberg:2012mz,Costa:2011mg,Costa:2011dw,SimmonsDuffin:2012uy,Kravchuk:2016qvl,Hogervorst:2013sma,Kravchuk:2017dzd,Karateev:2017jgd,Penedones:2015aga,Costa:2016hju,Costa:2016xah,Iliesiu:2015akf,Elkhidir:2014woa,Echeverri:2015rwa,Echeverri:2016dun,Cuomo:2017wme,Dymarsky:2013wla} in $d\geq3$ dimensions.
Impressive numerical methods~\cite{
Caracciolo:2009bx,Rychkov:2009ij,Rattazzi:2010gj,Poland:2010wg,Rattazzi:2010yc,Vichi:2011ux,Poland:2011ey,
Rychkov:2011et,ElShowk:2012hu,Gliozzi:2013ysa,Kos:2013tga,
Nakayama:2014lva,Caracciolo:2014cxa,Nakayama:2014sba,Paulos:2014vya,
Kos:2015mba,Iliesiu:2015qra,Kim:2015oca,Chester:2016wrc,Behan:2016dtz,Nakayama:2016knq,El-Showk:2016mxr,
ElShowk:2012ht,El-Showk:2014dwa,Kos:2014bka,Simmons-Duffin:2015qma,Kos:2016ysd,Dymarsky:2017xzb,Dymarsky:2017yzx}
further allowed to put constraints on dynamical (model dependent) quantities. 

In parallel to numerical developments, an analytic conformal bootstrap  method has been proposed~\cite{Fitzpatrick:2012yx,Komargodski:2012ek}. It allows to gain access to large-spin sector of the CFT spectrum. More precisely it was shown that for every pair of scalars $\phi_1$ and $\phi_2$ in the CFT spectrum with twists\footnote{See~\eqref{eq: twist definition} for precise definition of twist in 4D.} $\tau_1$ and $\tau_2$, there exist an infinite tower of traceless symmetric operators $\OO$ spanned by a non-negative integer $n$, whose twist has the form
\begin{equation}\label{eq:the_result}
\tau=\left(\tau_1+\tau_2+2n\right)+\mathcal O \left(1/\ell \right).
\end{equation}
These operators are often called the double-twist operators. Their OPE coefficients at the leading order reproduce the ones of generalized free theories (GFTs).\footnote{\label{foot:definition_GFT}
Generalized free theory, also known as mean field theory, is a very special type of CFT. It can be defined by a set of primaries called fundamental operators.
All the other primaries, called composite (multi-twist) operators, are constructed out of the fundamental ones and their derivatives. We refer to the composits made out two primaries as double-twist operators. 
Correlation functions in GFTs are computed using Wick contractions, which is equivalent to factorizing a given $n$-point function into 2-point functions in all possible ways. Contrary to free theories, GFT operators do not satisfy equations of motion.} 
The result~\eqref{eq:the_result} was obtained by studying the s-t bootstrap equations for the scalar 4-point function
\begin{equation}\label{eq:scalar_correlators}
\langle\phi_1\phi_1\phi_2\phi_2\rangle.
\end{equation}
The analytic techniques of~\cite{Fitzpatrick:2012yx,Komargodski:2012ek} were further developed in~\cite{Kaviraj:2015cxa,Alday:2015eya,Kaviraj:2015xsa,Alday:2015ota,Alday:2015ewa,Dey:2016zbg,Alday:2016njk,Alday:2016jfr,Li:2015rfa,Simmons-Duffin:2016wlq,Qiao:2017xif}.
The most recent progress has been made in~\cite{Caron-Huot:2017vep, Simmons-Duffin:2017nub}. Other closely related works are ~\cite{Hogervorst:2017kbj,Hogervorst:2017sfd,Gadde:2017sjg} and~\cite{Hartman:2015lfa,Hartman:2016dxc,Hartman:2016lgu}.

The obtained CFT result can be interpreted in AdS as a system of scalar particles with large relative angular momentum $\ell$. One can compute shifts of their binding energies due to gravity~\cite{Fitzpatrick:2014vua,Cornalba:2006xm,Cornalba:2006xk,Cornalba:2007zb}
and other interraction~\cite{Fitzpatrick:2010zm}.
This matches nicely the sub-leading correction in~\eqref{eq:the_result} governed by the energy-momentum tensor on the CFT side.

Extending the analytic techniques to correlators with spin is crucial for studying generic CFT data. The works~\cite{Li:2015itl,Hofman:2016awc} addressed correlators with two scalars and two traceless symmetric operators\footnote{In~\cite{Li:2015itl} the correlator $\langle J_\mu\, J_\nu\, J_\rho\, J_\sigma\rangle$ was also studied.} such as conserved currents $J_\mu$ and the energy-momentum tensor $T_{\mu\nu}$. This has allowed to prove~\cite{Hofman:2016awc} the conjectured conformal collider bounds~\cite{Hofman:2008ar}.
The work~\cite{Li:2015itl} deals solely with 3D CFTs, whereas in~\cite{Hofman:2016awc} the authors work in $d\geq 3$ dimensions. Analytic studies of fermions also got  some attention recently~\cite{vanLoon:2017xlq,Giombi:2017rhm}.

\paragraph{Goal of the paper}
In this work we will discuss the analytic method of~\cite{Fitzpatrick:2012yx,Komargodski:2012ek} in the context of generic spinning correlators in 4D~\cite{Elkhidir:2014woa,Echeverri:2015rwa,Echeverri:2016dun,Cuomo:2017wme}.\footnote{Other approaches to spinning correlators applicable to 4D were used in~\cite{Rejon-Barrera:2015bpa,Costa:2016hju,Kravchuk:2017dzd}.} We will follow the notation of~\cite{Cuomo:2017wme}. 
For performing computations in practice we will develop and attach to the paper a little Mathematica~\cite{Mathematica} code ``analyticBootstrap4D.m'', which should be used together with the ``CFTs4D'' package~\cite{Cuomo:2017wme}. We mention the names of relevant functions throughout the text as \fref{functionName}.

As a demonstration in this paper we will study in detail the simplest spinning correlator in 4D, namely the scalar-fermion correlator\footnote{This correlator is not unitary (or reflection-positive in the Euclidean signature).}
\begin{equation}\label{eq:demonstration}
\langle \phi\,
\psi\,
\phi\,
\overline\psi\rangle.
\end{equation}
We will address both s-t and u-t channels and discuss how they complement each other.\footnote{In a similar way one could study the u-t bootstrap equation of~\eqref{eq:scalar_correlators}, see~\cite{Li:2015rfa}.}
To facilitate the discussion we will also study the spectrum and the OPE coefficients of the scaclar-fermion GFT. We will discuss how the GFT results are related to the large-spin analytic results.
We will provide details of all the computations in the attached Mathematica notebook "Example.nb".

\paragraph{Outline of the paper}
In section~\ref{sec:analytic_bootstrap} we discuss in details generic spinning 4-point functions. We provide a recipe for studying them analytically and develop practical tools for performing computations in Mathematica.
In section~\ref{sec:bootstrap_equations} we consider a demonstration of the strategy given in section~\ref{sec:analytic_bootstrap} on the scalar-fermion 4-point function. A compact summary of the final result is given in section~\ref{sec:summary}.
In section~\ref{sec:GFT} we address a scalar-fermion GFT and show how it is related to our analytic bootstrap results. 
We conclude in section~\ref{sec:discussion}.
Details on the Ward identities for the scalar-fermion correlator are provided in appendix~\ref{sec:Ward_identities}.

\section{Analytic Bootstrap}
\label{sec:analytic_bootstrap}
We start by discussing the main ingredients of general bootstrap equations in section~\ref{sec:bootstrap_eq} preparing the ground for section~\ref{sec:recipe}, where we provide an analytic bootstrap recipe in a schematic form.

\subsection{Bootstrap Equations}
\label{sec:bootstrap_eq}
We work in 4D Minkowski space where the conformal group is $SO(2,4)\simeq SU(2,2)$. The local primary operators transform in a finite-dimensional representation of the $U(1)\times SU(2)\times SU(2)$ sub-group and thus are labeled by the $U(1)$ charge (the scaling dimension) $\Delta$ and the pair of integers $(\ell,\bar{\ell})$ which describes spin.\footnote{Traceless symmetric operators in 4D are those with $\ell=\bar{\ell}$.} In what follows it will be more convenient 
to use instead of the scaling dimension another quantity called the twist defined as
\be\label{eq: twist definition}
\tau\equiv \Delta-\frac{\ell+\ol\ell}{2},\quad \text{in unitary CFTs\;}
\begin{cases}
	\tau   \geq 1,       & \quad \ell\bar{\ell}=0,\\
	\tau   \geq 2,       & \quad \ell\bar{\ell}\neq 0.\\
\end{cases}
\ee
The inequalities in~\eqref{eq: twist definition} are the unitary bounds, they are saturated by conserved operators~\cite{Mack:1975je,Minwalla:1997ka}. In the $\ell\bar{\ell}=0$ case the conserved operators are free~\cite{Weinberg:2012cd}. A generic local operator is denoted by 
\begin{equation}\label{eq:generic_local_operators}
\OO^{(\ell,\bar{\ell})}_{k,\tau}(\point_i),\quad
\point_i\equiv (x_i,s_i,\bar s_i),
\end{equation}
where $x$ is the position and $s$ and $\bar s$ are auxiliary spinors (polarizations). The additional index $k$ is used to label degenerate operators, which arise for example in the case of global symmetries or Dirac fermions. In the reminder of this paper we drop this label for simplicity. We use a convention for naming operators~\eqref{eq:generic_local_operators} such that 
\begin{equation}\label{eq:convention}
\ell\geq\bar{\ell}.
\end{equation}
Every operator~\eqref{eq:generic_local_operators} has its hermitian conjugate
\begin{equation}
\ol\OO^{(\bar\ell,{\ell})}_\tau(\point_i)\equiv
\left(
\OO_\tau^{(\ell,\bar{\ell})}(\point_i)
\right)^\dagger.
\end{equation}

A generic 4-point function of local operators has the following form when expanded in the s-channel
\begin{equation}\label{eq:s-channel_decomposition}
\contraction{\langle}{\OO}{_1(\point_1)\,}{\OO}
\contraction{\langle \OO_1(\point_1)\,\OO_2(\point_2)\,}{\OO}{_3(\point_3)\,}{\OO}
\langle \OO_1(\point_1)\,\OO_2(\point_2)\,\OO_3(\point_3)\,\OO_4(\point_4) \rangle=
\sum_{\OO} \sum_{a,b} \lambda_{\langle\OO_1\OO_2\OO\rangle}^a \lambda_{\langle\ol\OO\OO_3\OO_4\rangle}^b \; W^{ab}_{\langle\OO_1\OO_2\OO\rangle\langle\ol\OO\OO_3\OO_4\rangle},
\end{equation}
where $\lambda^a$ are the OPE coefficients (3-point coupling constants) and $W^{ab}$ are the conformal partial waves (CPWs) which can be written as
\begin{equation}
W^{ab}=\sum_{I} H_I^{ab}(z,\zb)\; \mathbb{T}^I,
\end{equation}
where $H$ are conformal blocks and  $\mathbb{T}^I$ are the tensor structures of the correlator~\eqref{eq:s-channel_decomposition}.
We will refer to the operators $\OO_i$ in correlators as external operators and to the operator $\OO$ as internal (exchanged). The sum over all possible operators should be seen as a sum over all possible values of twist and spin
\begin{equation}\label{eq:summation}
\sum_{\OO}=\sum_{\tau}\sum_{\ell,\,\bar\ell}.
\end{equation}
Any conformal block $H_I^{ab}(z,\zb)$ can be expressed~\cite{Echeverri:2015rwa} in terms of the derivatives of simpler blocks called the seed blocks and dual seed blocks.
Both seed and dual seed blocks have the following form~\cite{Echeverri:2016dun}
\begin{equation}\label{eq:schematic_expression_seed_blocks}
H_{seed}^{(p)}(z,\bar z)=
\Big(\frac{z \bar z}{z-\bar z}\Big)^{2\,p+1}\, \sum_{m,n}c_{m,n}\mathcal{F}_{\rho_1\,\rho_2}^{(a,b;c)}(z,\bar z),
\quad p\equiv|\ell-\bar{\ell}|.
\end{equation}
The coefficients $c_{m,n}$ are some constants~\cite{Echeverri:2016dun,Cuomo:2017wme} which have the following schematic form
\begin{equation}\label{eq:Coefficients}
c_{m,n}^e = (-1)^\ell\;i^p\times rationalFunction_{m,n}(a,b;\;\ell,\tau).
\end{equation}
An example of how the $rationalFunction$ looks like can be found in appendix B in~\cite{Echeverri:2016dun} for the $p=1$ case.
The functions $\mathcal{F}$ are defined as
\begin{equation}\label{eq:definition_F-function}
\mathcal{F}^{(a,b;c)}_{\rho_1\rho_2}(z,\zb)\equiv k_{\rho_1}^{(a,b;c)}(z)k_{\rho_2}^{(a,b;c)}(\bar z)- (z \leftrightarrow \bar z).
\end{equation}
Here the parameters $a$ and $b$ depend on the twists of external operators, parameter $c$ on the spin difference $p$ defined in~\eqref{eq:schematic_expression_seed_blocks}. The parameters $\rho_1$ and $\rho_2$ depend on twist and spin of the internal (exchanged) operator as
\begin{equation}\label{eq:definitions_of_parameters_rho}
\rho_1\equiv\frac{\tau}{2}+l+r,\quad
\rho_2\equiv\frac{\tau}{2}+s,
\end{equation}
where $r$ and $s$ are some integer or half- integer constants. See~\cite{Echeverri:2016dun} for details.
The $k$-functions are defined as
\begin{equation}\label{eq:definition_k-function}
k_\rho^{(a,b;c)}(x)\equiv x^\rho{}_2F_1(a+\rho,\,b+\rho;\,c+2\rho;\,x).
\end{equation}
For $b=c-a$ the k-functions admit the following representation as a double-infinite sum
\begin{multline}\label{eq:k_expansion}
k_{\rho}^{(a,\,c-a;\,c)}(1-x)=\frac{\Gamma(2\rho+c)}{\Gamma(\rho+a)\Gamma(\rho-a+c)}\sum_{m=0}^{\infty}\sum_{n=0}^{\infty}
\frac{(-\rho)_m}{m!}\frac{(\rho+a)_n}{n!}\frac{(\rho+c-a)_n}{n!}\,x^{m+n}\\
\times\Big(2\psi(n+1)-\psi(\rho+a+n)-\psi(\rho+c-a+n)-\ln(x)\Big),\quad x\in[0,1],
\end{multline}
which simply follows from Taylor expansion around $x=0$. The Pochhammer symbol and the digamma functions are denoted by
\begin{equation}
(a)_n   \equiv \frac{\gammafn{a+n}}{\gammafn{a}},\qquad
\psi(b) \equiv \frac{\Gamma^\prime(b)}{\gammafn{b}}
\end{equation}
respectively. Another important representation of~\eqref{eq:definition_k-function} can be obtained in the large $\ell>>1$ limit and small $x<<1$, which reads as
\begin{equation}\label{eq:hypergeom-into-bessel}
\lim_{\ell\rightarrow \infty}
\lim_{x\rightarrow 0}\;
k_{\rho_1}^{(a,b;c)}(1-x) = 
\sqrt{\frac{\ell}{\pi}}\,2^{2\rho_1+c}x^{-\frac{a+b-c}{2}}K_{a+b-c}(2\ell\sqrt{x}),\quad\ell^2 x\sim 1,
\end{equation}
where $K_\alpha$ is the modified Bessel function and the parameter $\rho_1$ is defined in~\eqref{eq:definitions_of_parameters_rho}.\footnote{Notice that~\eqref{eq:hypergeom-into-bessel} contains a factor $4^\ell$ entering trough the parameter $\rho_1$.} This relation is a straightforward generalization of the results of appendix A.1.4 in~\cite{Fitzpatrick:2012yx}.

For a given correlator one can write essentially 2 bootstrap equations: the s-t equation
\begin{equation}\label{eq:s-t-bootstrap_equations}
\contraction{\langle}{\OO}{_1(\point_1)\,}{\OO}
\contraction{\langle \OO_1(\point_1)\,\OO_2(\point_2)\,}{\OO}{_3(\point_3)\,}{\OO}
\langle \OO_1(\point_1)\,\OO_2(\point_2)\,\OO_3(\point_3)\,\OO_4(\point_4) \rangle
=
\contraction{\langle}{\OO}{_1(\point_1)\,\OO_2(\point_2)\,\OO_3(\point_3)\,}{\OO}
\contraction[2ex]{\langle\OO_1(\point_1)\,}{\OO}{_2(\point_2)\,}{\OO}
\langle \OO_1(\point_1)\,\OO_2(\point_2)\,\OO_3(\point_3)\,\OO_4(\point_4) \rangle
\end{equation}
and the u-t equation
\begin{equation}\label{eq:u-t-bootstrap_equations}
\contraction{\langle}{\OO}{_1(\point_1)\,\OO_2(\point_2)\,}{\OO}
\contraction[2ex]{\langle \OO_1(\point_1)\,}{\OO}{_2(\point_2)\,\OO_3(\point_3)\,}{\OO}
\langle \OO_1(\point_1)\,\OO_2(\point_2)\,\OO_3(\point_3)\,\OO_4(\point_4) \rangle
=
\contraction{\langle}{\OO}{_1(\point_1)\,\OO_2(\point_2)\,\OO_3(\point_3)\,}{\OO}
\contraction[2ex]{\langle\OO_1(\point_1)\,}{\OO}{_2(\point_2)\,}{\OO}
\langle \OO_1(\point_1)\,\OO_2(\point_2)\,\OO_3(\point_3)\,\OO_4(\point_4) \rangle.
\end{equation}

\paragraph{s-t bootstrap equation}
We consider now the bootstrap equation~\eqref{eq:s-t-bootstrap_equations}. After decomposing it into a set of independent equations by stripping of independent tensor structures $\mathbb{T}^I$, all the equations get the following schematic form
\begin{equation}\label{eq:typical_bootstrap_equation}
(z \zb)^A\sum_{\OO} P_{\OO} \; H_{\OO}^I(z,\zb)=
\Big((1-z)(1-\zb)\Big)^A\sum_{\OO} P'_{\OO}\; H^{\prime\;I}_{\OO}(1-z,1-\zb).
\end{equation}
It can be studied in the $z,\zb\in[0,1]$ region. Here $P_{\OO}$ and $P'_{\OO}$ represent products of the OPE coefficients
\begin{equation}
P_{\OO}\equiv\lambda_{\langle\OO_1\OO_2\OO\rangle}^a \lambda_{\langle\ol\OO\OO_3\OO_4\rangle}^b,\quad
P'_{\OO}\equiv\lambda_{\langle\OO_3\OO_2\OO\rangle}^a \lambda_{\langle\ol\OO\OO_1\OO_4\rangle}^b
\end{equation}
and $H^I_{\OO}$, $H^{\prime\;I}_{\OO}$  represent the conformal blocks which can be constructred from the (dual) seed blocks~\eqref{eq:schematic_expression_seed_blocks} and their derivatives.
The multipliers $(z \zb)^A$ and $\Big((1-z)(1-\zb)\Big)^A$ come from the ratio of kinematic factors in the s- and t-channel with the exponent $A$ which depends on twist of external operators.

\begin{figure}[t]
	\label{fig:complex_plain}
\centering
\includegraphics[scale=0.5]{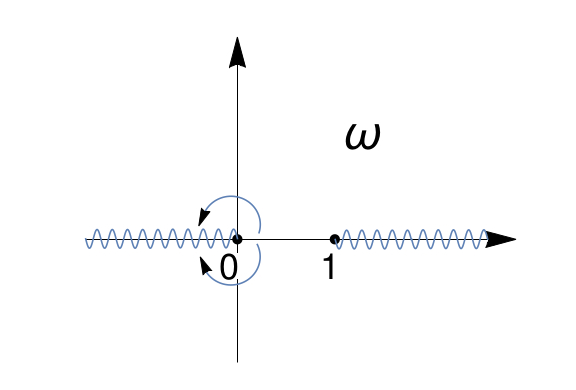}
\caption{Complex plain $w$. There are two branch cuts for the $k$ function: one is due to the hypergeometric series and another one due to the power-law behavior. By circular arrows we show 2 different ways to perform an analytic continuation from the $\omega \in[0,1]$ to the $[-\infty,0]$ line.}
\end{figure}

\paragraph{u-t bootstrap equation}
Equation~\eqref{eq:u-t-bootstrap_equations} leads to the following set of bootstrap equations
\begin{equation}\label{eq:u-t-equation_schematic}
\left(\frac{1}{z\,\zb}\right)^A\sum_{\OO} P''_{\OO} \; H^{\prime\prime\;I}_{\OO}\left(\frac{1}{z},\,\frac{1}{\zb}\right)=
\Big((1-z)(1-\zb)\Big)^A\sum_{\OO} P'_{\OO}\; H^{\prime\;I}_{\OO}(1-z,1-\zb).
\end{equation}
We will study this equation in the $z,\zb\in[1,+\infty]$ region. We make a change of variables from $(z,\bar z)$ to $(\omega,\omegab)$ defined as 
\be\label{eq:omega_variables}
\omega\equiv\frac{1}{z},\quad \omegab\equiv\frac{1}{\zb},\quad
\omega,\,\omegab\in[0,1].
\ee
Even though the hypergeometric series defining the $k(\omega)$ functions~\eqref{eq:definition_k-function} is convergent in the $\omega\in[0,1]$ region, the $k(\omega)$ functions is analytic in the whole complex plane modulo branch cuts~\cite{Rychkov:2017tpc}, see figure~\ref{fig:complex_plain}. The following formula holds~\cite{Tables}
\be\label{eq:analytic_continuation_formula_k}
k^{(a,b;c)}_{\rho}\left(1-\frac{1}{\omega}\right) = (-1)^{\pm\rho}\left(\omega\right)^a\; k^{(a,c-b;c)}_{\rho}(1-\omega).
\ee
The factor $(-1)^{\pm\rho}$ appears due to the choice of analytic continuation $\pm i\pi\epsilon$, namely one can choose to be right above or right below the left brunch cut, as indicated in figure~\ref{fig:complex_plain} by arrows.
We perform an analytic continuation of the functions $k(\omega)$ and $k(\omegab)$ separately choosing opposite directions. It is required when moving to Euclidean space where $\omegab=\omega^*$. As a consequence we get
\be\label{eq:analytic_continuation_formula_F}
\mathcal F^{(a,b;c)}_{\rho_1,\rho_2}\left(1-\frac{1}{\omega},1-\frac{1}{\omegab}\right) = (-1)^{\rho_1-\rho_2}\left(\omega \omegab\right)^a\; \mathcal F^{(a,c-b;c)}_{\rho_1,\rho_2}(1-\omega,1-\omegab),
\ee
where according to~\eqref{eq:definitions_of_parameters_rho} $\rho_1-\rho_2$ is an integer and one has
\begin{equation}\label{eq:minusSignAnalyticContinuation}
(-1)^{\rho_1-\rho_2}=\pm(-1)^{\ell}.
\end{equation}
After these manipulations instead of equation~\eqref{eq:u-t-equation_schematic} we can write
\begin{equation}\label{eq:u-t-equation_schematic_new}
\left(\omega \omegab\right)^{2A-a}\sum_{\OO} P''_{\OO} \; H^{\prime\prime\;I}_{\OO}(\omega,\,\omegab)=
\Big((1-\omega)(1-\omegab)\Big)^A\sum_{\OO} P'_{\OO}\;(-1)^\ell\; H^{\prime\prime\prime\;I}_{\OO}(1-\omega,1-\omegab).
\end{equation}
Structurally it has exactly the same form as~\eqref{eq:typical_bootstrap_equation} but with an extra $(-1)^\ell$.

There is an equivalent and somewhat simpler way of dealing with the u-t bootstrap equations. One can consider a different correlation function (with positions of operators re-ordered) and to study an s-t channel
\begin{equation}\label{eq:alternative_approach_u-t_equations}
\contraction{\langle}{\OO}{_1(\point_1)\,}{\OO}
\contraction{\langle \OO_1(\point_1)\,\OO_2(\point_2)\,}{\OO}{_3(\point_3)\,}{\OO}
\langle \OO_1(\point_1)\,\OO_3(\point_2)\,\OO_2(\point_3)\,\OO_4(\point_4) \rangle
=
\contraction{\langle}{\OO}{_1(\point_1)\,\OO_2(\point_2)\,\OO_3(\point_3)\,}{\OO}
\contraction[2ex]{\langle\OO_1(\point_1)\,}{\OO}{_2(\point_2)\,}{\OO}
\langle \OO_1(\point_1)\,\OO_3(\point_2)\,\OO_2(\point_3)\,\OO_4(\point_4) \rangle.
\end{equation}
In this case the standard s-t techniques straightforwardly apply and no analytic continuation is needed.

\subsection{The Analytic Bootstrap Recipe}
\label{sec:recipe}
In Minkowski space the variables $z$ and $\zb$ are independent quantities. The light-cone limit is defined as\footnote{In the $\omega$-variable we will use the light-cone limit $\omega\ll1-\overline\omega\ll1$.}
\begin{equation}\label{eq:light-cone_limit}
z\ll1-\zb\ll1.
\end{equation}
We summarize here the recipe of~\cite{Fitzpatrick:2012yx} for studying analytically the bootstrap equations of the form~\eqref{eq:typical_bootstrap_equation} in the light-cone limit~\eqref{eq:light-cone_limit}.

We keep the discussion at the schematic level and pretend that the conformal blocks $H$ are simply represented by a single $\mathcal F$ function~\eqref{eq:definition_F-function} together with the $(-1)^\ell\,i^p$ factor coming from~\eqref{eq:Coefficients}\footnote{We thus ignore the necessity of taking a linear combination in $F$ with different coefficients $c_{m,n}$ and taking derivatives in $z$ and $\zb$ in general situation. We also drop the factor $\Big(\frac{z \bar z}{z-\bar z}\Big)^{2\,p+1}$  which in the limit~\eqref{eq:light-cone_limit} simply leads to $(-z)^{2\,p+1}$.}
\begin{equation}\label{eq:toyBlock}
H(z,\bar z)=
 (-1)^\ell\;i^p\times\mathcal{F}_{\rho_1\,\rho_2}^{(a,b;c)}(z,\bar z).
\end{equation}

\paragraph{The left-hand side of~\eqref{eq:typical_bootstrap_equation}}  in the light-cone limit reads as\footnote{In writing~\eqref{eq:lhs} we assume $\rho_1<\rho_2$. Depending on parameters $r$ and $s$ defined in~\eqref{eq:definitions_of_parameters_rho} there might be a situation when $\rho_1>\rho_2$. The second term in~\eqref{eq:lhs} should change accordingly.} \fref{KeepLeadingTerm}
\begin{equation}\label{eq:lhs}
(z\zb)^{A}\left(1 + i^p\sum_{\ell_{min}} (-1)^{\ell_{min}} \sum_{\tau_{min}>1}^{2} z^{\rho_1}\; k_{\rho_2}^{(a,b;c)}(\bar z)\right),\quad A<0.
\end{equation}
The first term comes from the identity operator that may or may not be present. This depends on the correlator under consideration. The second term describes the contribution of all the rest of operators in the theory. In the $z\rightarrow 0$ limit only the lowest (minimal) twist operators contribute (light scalars or fermions with $1<\tau\leq2$ and conserved bosonic or fermionic operators with $\tau=2$).
We then recast the expression~\eqref{eq:lhs} into the final form by 
performing a series expansion around small $1-\zb$ \fref{kFunctionExpand}.\footnote{
The action of \fref{kFunctionExpand} on $k(\zb)$ is equivalent to~\eqref{eq:k_expansion} for $x=1-\zb$ and $b=c-a$.}

\paragraph{The right-hand side of~\eqref{eq:typical_bootstrap_equation}} will be dominated (in the light-cone limit) by the large spin operators as was argued in~\cite{Fitzpatrick:2012yx,Komargodski:2012ek}.\footnote{The poles in~\eqref{eq:lhs} for $z\rightarrow 0$ can only be reproduced by an infinite sum over large spin.} Using the definition~\eqref{eq:definitions_of_parameters_rho} it is clear that the second term in~\eqref{eq:definition_F-function} is negligible since it behaves as $(1-\zb)^{\ell}$ and should be dropped using \fref{plugSeedBlocks[0], plugDualSeedBlocks[0]}. We approximate the sum over $\ell$ by an integral
\begin{equation}
\sum_{\ell}=\int_0^\infty d\ell\;\mathcal{R}_\tau(\ell),
\quad P_\OO=P_{\tau}(\ell),
\end{equation}
where the distribution density $\mathcal{R}_\tau(\ell)$ indicates how operators with a given twist $\tau$ contribute to the sum.
The right-hand side of~\eqref{eq:typical_bootstrap_equation} is then written as\footnote{Notice that the equation~\eqref{eq:rhs} would have not contained the factor $(-1)^\ell$ if we had started from the bootstrap equation~\eqref{eq:u-t-equation_schematic_new}.}
\fref{largeSpinExpand}
\begin{multline}\label{eq:rhs}
2^{2t+c}\,z^{-\frac{a+b-c}{2}}\times \sum_{\tau}
(1-\zb)^A\, 2^{\tau}\;k_{\rho_2}^{(a,b;c)}(1-\zb)\\
\times\int_0^{\infty}d\ell\;\rho_\tau(\ell)\;P_\tau(\ell)\;(-1)^\ell\;i^p\;4^\ell\;\sqrt{\frac{\ell}{\pi}}\;K_{a+b-c}(2\ell\sqrt{z}).
\end{multline}

In the large $\ell$ limit all the coefficients~\eqref{eq:Coefficients} simplify significantly but remain non-zero for generic parameters. They do not give any $\ell$ contribution apart from the $(-1)^\ell$ factor. This is the reason we can safely leave them out of the discussion in~\eqref{eq:toyBlock}.

In the large spin limit we expect the following behavior of the squared OPE coefficients multiplied by the density
\begin{equation}\label{eq:OPE_coefficients_behavior}
\mathcal{R}_\tau(\ell)\; P_\tau(\ell)=4^{-\ell}\;\ell^B\;\left(P^0+\frac{P^1}{\ell^C}+\ldots\right),
\end{equation}
where the coefficients $P^0$ and $P^1$ contain the dependence on $\ell$ only via the factor $(-1)^\ell$ which may or may not be present. In what follows we assume
$\mathcal{R}_\tau(\ell)=1$.\footnote{This hypothesis seems to be confirmed for the 3D Ising model in~\cite{Simmons-Duffin:2016wlq}.} Presence of the $4^{-\ell}$ factor in~\eqref{eq:OPE_coefficients_behavior} guaranties sufficiently fast decay of the OPE coefficients required by the OPE convergence~\cite{Pappadopulo:2012jk,Rychkov:2015lca} and should cancel out the $4^\ell$ factor in~\eqref{eq:rhs}. The same type of ansatz was also used in~\cite{Li:2015itl,Hofman:2016awc}.

Using~\eqref{eq:OPE_coefficients_behavior} one can then evaluate the integral in~\eqref{eq:rhs} analytically term by term \fref{largeSpinSum} by means of
\be\label{eq:the_integral}
\int_0^\infty d\ell \;\ell^\alpha\;K_\beta(2\ell\sqrt{z})=
\frac{1}{4}\,z^{-\frac{\alpha+1}{2}}\,
\gammafn{\frac{1+\alpha+\beta}{2}}\gammafn{\frac{1+\alpha-\beta}{2}}.
\ee
It is clear that, if the factor $(-1)^\ell$ is present in the final expression, the integral would return zero, since contributions from even and odd spins are approximately the same and cancel each other out.

The twist has the following behavior
\begin{equation}\label{eq:twist_behavior}
\tau=\tau^0+\frac{\gamma}{\ell^D}+\ldots,
\end{equation}
where $\tau^0$ is the leading twits and $\gamma$ is the coefficient of the sub-leading correction often called the anomalous dimension.
Plugging~\eqref{eq:twist_behavior} to~\eqref{eq:rhs} one should make an expansion around $\tau^0$ for small $\frac{\gamma}{\ell^D}$ of the expression
\begin{equation}
2^{\tau}\;k_{\rho_2}^{(a,b;c)}(1-\zb)\approx
2^{\tau^0}\;k_{\frac{\tau^0}{2}+s}^{(a,b;c)}(1-\zb)
+
\frac{\partial}{\partial \tau}\left(2^{\tau}\;k_{\frac{\tau}{2}+s}^{(a,b;c)}(1-\zb)\right) \bigg|_{\tau=\tau^0}\times\frac{\gamma}{\ell^D}.
\end{equation}
We then recast the expression~\eqref{eq:rhs} into the final form by 
performing a series expansion around small $1-\zb$ \fref{kFunctionExpand}. For example at the leading order in $1-\zb$ one has
\begin{equation}
2^{\tau_{\OO}}\;k_{\rho_2}^{(a,b;c)}(1-\zb)\approx
2^{\tau^0}(1-\zb)^{s+\frac{\tau^0}{2}}\times
\left(1+\frac{\gamma}{2\,l^D}\;\log\big(4(1-\zb)\big) \right).
\end{equation}

\paragraph{Matching the left- and right-hand side}
After performing all prescribed manipulations one can equating~\eqref{eq:lhs} and~\eqref{eq:rhs} and determine the unknown coefficients in~\eqref{eq:OPE_coefficients_behavior} and~\eqref{eq:twist_behavior} by matching independently terms with equal powers 
\begin{equation}
z^{\#_1}\times(1-\zb)^{\#_2}\times \log^{\#_3}(1-\zb),
\end{equation}
where $\#_3$ can only be $0$ or $1$. More precisely we start analyzing terms with $\#_3=0$. Matching the power $\#_1$ allows to determine the exponent $B$ and matching the power $\#_2$ allows to determine the leading twist $\tau^0$. We have a series expansion in $1-\zb$, to satisfy the equations order by order in $1-\zb$ it is required to have an infinite tower of operators with the twists $\tau^0$ which differ by $2n$, where $n$ is a non-negative integer. We can then determine the coefficient $P^0$.
By studying the terms with $\#_3=1$ one gets access to the anomalous dimension. After extracting the coefficient $\gamma$ and the exponent $D$, one can also compute $P^1$.

In the case of bootstrap equations with no identity operator we do not get access to all the parameters, we will see an instance of this in section~\ref{subsec:s-t}.

\section{Scalar-Fermion Correlator}
\label{sec:bootstrap_equations}
As the simplest example of spinning correlator we consider the following 4-point function
\begin{equation}\label{eq:the_correlator}
\langle \phi(\point_1)\,
\psi(\point_2)\,
\phi(\point_3)\,
\overline\psi(\point_4)\rangle,
\end{equation}
where $\phi$ is some real scalar\footnote{The case of complex scalar is almost identical to the case of real scalar. The only difference hides in the properties of the OPE coefficients.}, $\psi$ and $\bar{\psi}$ are a fermion and its hermitian conjugate in the representations $(1,0)$ and $(0,1)$ respectively.
We first study the structure of this correlator (tensor structures, decomposition into OPE coefficients and conformal blocks) providing all the necessary technical ingredients.
We then apply the machinery of section~\ref{sec:analytic_bootstrap} and study
the u-t and s-t bootstrap equations. We combine the results for both channels and show how they complement each other providing a compact summary in section~\ref{sec:summary}.

\subsection{Structure of the Correlator}
The correlator~\eqref{eq:the_correlator} has the following structure
\begin{equation}\label{eq:TheCorrelationFunction}
\langle \phi(\point_1)\,
\psi(\point_2)\,
\phi(\point_3)\,
\overline\psi(\point_4)\rangle=\sum_{I=1}^2 g^I(z,\bar z)\,\mathbb{T}^I,
\end{equation}
where $g^I(z,\bar z)$ is an arbitrary function of conformally invariant variables and $\mathbb{T}^I$ are the tensor structures defined as
\begin{equation}
\mathbb{T}^1\equiv \mathcal{K}_4\;\hat \II^{42},\quad
\mathbb{T}^2\equiv \mathcal{K}_4\;\hat \II^{42}_{31},\quad
\mathcal K_4 \equiv \left(x_{12}^2\;x_{34}^2\right)^{-\frac{\tau_\phi+\tau_\psi+1}{2}}\left(\frac{x_{13}^2}{x_{24}^2}\right)^{\frac{1+\tau_\psi-\tau_\phi}{2}}.
\end{equation}
The objects $\hat \II^{42}$ and $\hat \II^{42}_{31}$ represent the spin structure of the correlator, their explicit form is irrelevant for current discussion (see~\cite{Cuomo:2017wme} for precise definitions).

We address now the s-, t- and u-channel OPE decompositions of the correlator~\eqref{eq:TheCorrelationFunction}.
For brevity we will use the following names for bosonic (traceless symmetric) and fermionic operators consistent with~\eqref{eq:convention}
\begin{eqnarray}\label{eq:the_operator_set}
\OO\equiv \OO^{(\ell,\ell)}_{\tau}
\quad{\bf and}\quad
\Psi\equiv\OO^{(\ell+1,\ell)}_{\tau}.
\end{eqnarray}
It is convenient to write explicitly the contribution of operators and their hermitian conjugates to the sum~\eqref{eq:summation}. The decomposition~\eqref{eq:s-channel_decomposition} reads then 
\begin{align}
\label{eq:s-channel}
\contraction{\langle}{\phi}{(\point_1)\,}{\psi}
\contraction{\langle \phi(\point_1)\,\psi(\point_2)\,}{\psi}{(\point_3)\,}{\psi}
\langle \phi(\point_1)\,\psi(\point_2)&\,\phi(\point_3)\,\overline\psi(\point_4) \rangle=\nonumber\\
&=\sum_{\Psi}\Big(
\underbrace{\lambda_{\langle\phi\psi\Psi\rangle}\lambda_{\langle\overline\Psi\phi\overline\psi\rangle}}_{\equiv P_{\Psi\overline\Psi}}\,W^{dual\;seed}_{\langle\phi\psi\Psi\rangle\langle\overline\Psi\phi\overline\psi\rangle}+
\underbrace{\lambda_{\langle\phi\psi\overline\Psi\rangle}\lambda_{\langle\Psi\phi\overline\psi\rangle}}_{\equiv P_{\overline\Psi\Psi}}\,W^{seed}_{\langle\phi\psi\overline\Psi\rangle\langle\Psi\phi\overline\psi\rangle}
\Big),\\
\label{eq:t-channel}
\contraction{\langle}{\phi}{(\point_1)\,\psi(\point_2)\,\phi(\point_3)\,}{\psi}
\contraction[2ex]{\langle\phi(\point_1)\,}{\psi}{(\point_2)\,}{\phi}
\langle \phi(\point_1)\,\psi(\point_2)&\,\phi(\point_3)\,\overline\psi(\point_4) \rangle=\nonumber\\
&=\sum_{\Psi}\Big(
\underbrace{\lambda_{\langle\phi\psi\Psi\rangle}\lambda_{\langle\overline\Psi\phi\overline\psi\rangle}}_{\equiv P_{\Psi\overline\Psi}}\,W^{dual\;seed}_{\langle\phi\psi\Psi\rangle\langle\overline\Psi\phi\overline\psi\rangle}+
\underbrace{\lambda_{\langle\phi\psi\overline\Psi\rangle}\lambda_{\langle\Psi\phi\overline\psi\rangle}}_{\equiv P_{\overline\Psi\Psi}}\,W^{seed}_{\langle\phi\psi\overline\Psi\rangle\langle\Psi\phi\overline\psi\rangle}
\Big)
\Bigg|_{\point_1\leftrightarrow \point_3},\\
\label{eq:u-channel}
\contraction{\langle}{\phi}{(\point_1)\,\psi(\point_2)\,}{\phi}
\contraction[2ex]{\langle \phi(\point_1)\,}{\psi}{(\point_2)\,\phi(\point_3)\,\overline}{\psi}
\langle \phi(\point_1)\,\psi(\point_2)&\,\phi(\point_3)\,\overline\psi(\point_4) \rangle=\nonumber\\
&=\langle \phi(\point_1)\phi(\point_3)\,\rangle\;\langle\psi(\point_2)\,\overline\psi(\point_4) \rangle+\sum_{\OO}\underbrace{\lambda_{\langle\phi\phi\OO\rangle}\lambda_{\langle\OO\psi\overline\psi\rangle}^a}_{\equiv P_{\OO}^a}\,W_{\langle\phi\phi\OO\rangle\langle\OO\psi\overline\psi\rangle}^a\Bigg|_{\point_2\leftrightarrow \point_3}.
\end{align}
The labels $seed$ and $dual\;seed$ indicate that the CPWs have the simplest possible structure among the spinning blocks in 4D, see~\eqref{eq:schematic_expression_seed_blocks}. Here we work with $p=1$ (dual) seeds. In the last equation the CPWs $W^a$ can be computed from the seed CPW with $p=0$. We provide below all the relevant information on  how to do it.

The products of OPE coefficients $\lambda$ are generically denoted by $P$. The OPE coefficients $\lambda$ are defined via the basis of tensor structures in a given 3-point function. We choose the following basis of structures for correlators involving $\Psi$ and $\ol\Psi$
\begin{align}\label{eq:def_1}
\langle \phi(\point_1)\,\psi(\point_2)\,\overline\Psi(\point_3)\rangle &=
\lambda_{\langle\phi\psi\overline\Psi\rangle} \,\mathcal{K}_3\, \hat \II^{32} [\hat \JJ^3_{12}]^{\ell-1},\\\label{eq:def_2}
\langle \phi(\point_1)\,\psi(\point_2)\,\Psi(\point_3)\rangle &=
\lambda_{\langle\phi\psi\Psi\rangle} \,\mathcal{K}_3\, \hat \KK^{23}_1 [\hat \JJ^3_{12}]^{\ell-1},\\\label{eq:def_3}
\langle\overline\Psi(\point_1)\,\phi(\point_2)\,\overline\psi(\point_3)\,\rangle &=
\lambda_{\langle\overline\Psi\phi\overline\psi\rangle} \,\mathcal{K}_3\, \hat{\overline{\KK}}^{13}_2 [\hat \JJ^1_{23}]^{\ell-1},\\\label{eq:def_4}
\langle \Psi(\point_1)\,\phi(\point_2)\,\overline\psi(\point_3)\,\rangle &=
\lambda_{\langle\Psi\phi\overline\psi\rangle} \,\mathcal{K}_3\, \hat \II^{31} [\hat \JJ^1_{23}]^{\ell-1}
\end{align}
and for correlators involving $\OO$
\begin{align}\label{eq:def_5}
\langle \phi(\point_1)\,\phi(\point_2)\,\OO(\point_3)\rangle &=
\lambda_{\langle\phi\phi\OO\rangle} \,\mathcal{K}_3\, [\hat \JJ^3_{12}]^{\ell-1},\\
\label{eq:def_6}
\langle \OO(\point_1)\psi(\point_2)\,\bar\psi(\point_3)\rangle &=
\mathcal{K}_3 [\hat \JJ^1_{23}]^{\ell-1}\, \Brk{\lambda_{\langle\OO\psi\bar\psi\rangle} ^1 \hat\II^{32}\hat \JJ^1_{23} +\lambda_{\langle\OO\psi\bar\psi\rangle}^2\hat\II^{12}\hat\II^{31}}.
\end{align}
The object $\mathcal K_3$, $\hat\II^{ij}$, $\hat \JJ^i_{jk}$, $\hat{\overline{\KK}}^{jk}_i $ and $\hat{\KK}^{jk}_i $ encode spin and scaling structure of correlators. Their explicit form is irrelevant for current discussion (see~\cite{Cuomo:2017wme} for precise definitions).

The four OPE coefficients defined via~\eqref{eq:def_1}-\eqref{eq:def_4} are not independent of each other. Using complex conjugation of the defining correlators and permuting operators one gets the following relations\footnote{Since the products of OPE coefficients form positive (negative) definite quantities one can try to study the s-t bootstrap equations numerically using standard techniques. However since s- and t- channels are not unitary (or reflection positive) one does not expect to have any bounds. This is because (as we will see later) for such configuration the correlation function has an alternating factor $(-1)^\ell$ which prevents one from constructing a positive definite functional.}
\begin{equation}\label{eq:positivity_of_OPE_coefficients}
\begin{cases}
\lambda_{\langle\phi\psi\Psi\rangle}^*=
\lambda_{\langle\overline\Psi\phi\overline\psi\rangle}\\
\lambda_{\langle\phi\psi\overline\Psi\rangle}^*=-
\lambda_{\langle\Psi\phi\overline\psi\rangle}
\end{cases}
\Longrightarrow\quad
\begin{cases}
P_{\Psi\overline\Psi}=\big|\lambda_{\langle\phi\psi\Psi\rangle}\big|^2\\
P_{\overline\Psi\Psi}=-\big|\lambda_{\langle\phi\psi\overline\Psi\rangle}\big|^2.
\end{cases}
\end{equation}
Further constraints on the scaling dimensions and the OPE coefficients of external operators in~\eqref{eq:def_1}-\eqref{eq:def_6} appear if operators $\Psi$ and $\OO$ are conserved. 

In case $\OO$ is conserved, the correlators~\eqref{eq:def_5}, \eqref{eq:def_6} automatically satisfy the differential conservation equations. The only further constraint comes from the Ward identities discussed in appendix~\ref{sec:Ward_identities} which allow to relate the OPE coefficients to normalization constant of 2-point functions. We use the normalization convention of~\cite{Cuomo:2017wme}.
In the case of the energy-momentum tensor $T$ we have
\begin{equation}\label{eq:wardIdentitiesStressTensor}
\lambda_{\langle\phi\phi T\rangle}=-\frac{2\tau_\phi}{3\pi^2},\quad
\lambda_{\langle T\psi\opsi\rangle}^{1}=\frac{2i\,(\tau_\psi-1)}{3 \pi^2},\quad
\lambda_{\langle T \psi\opsi\rangle}^{2}= -\frac{2i}{\pi^2}.
\end{equation}

In case $\Psi$ is conserved, we deal with supersymmetric current. Conservation condition implies that in the correlators~\eqref{eq:def_1}, \eqref{eq:def_4} and in~\eqref{eq:def_2}, \eqref{eq:def_3} the twists of external operators should be related as
\begin{equation}
\tau_{\phi}=\tau_{\psi}
\quad{\bf and}\quad
\tau_{\phi}=\tau_{\psi}+1
\end{equation}
respectively. This is expected in SUSY theories where scalars and fermions belong to the same multiplets. In principle we could have also used Ward identities to compute the OPE coefficients in~\eqref{eq:def_1}-\eqref{eq:def_4} in the presence of SUSY current, but this is beyond the scope of our work.

Finally we can construct tensor structures in~\eqref{eq:def_6} from the single scalar (seed) tensor structure by means of differential operators~\cite{Karateev:2017jgd}  
\begin{equation}\label{eq:differential_tensor_structures}
\langle \OO(\point_1)\psi(\point_2)\,\bar\psi(\point_3)\rangle=
\begin{pmatrix}
\lambda_{\langle\OO\psi\bar\psi\rangle} ^1, & \lambda_{\langle\OO\psi\bar\psi\rangle}^2
\end{pmatrix}
\times{\bf M}\times {\bf D}\;
\langle \OO(\point_1)\phi(\point_2)\,\phi(\point_3)\rangle^{(\bullet)},
\end{equation}
where $(\bullet)$ stresses that we do not deal with the full correlator but only with its kinematic part (tensor structure), the differential basis of operators ${\bf D}$ and the matrix of coefficients ${\bf M}$ can be chosen as follows
\begin{equation}\begin{aligned}\label{eq:differential_operators}
{\bf D}&=\begin{pmatrix}
\mathcal{D}^3_{-0+}\cdot\overline{\mathcal{D}}_{2}^{-+0}\Big|_{\tau_\phi\rightarrow\tau_\psi+1},\; & \;\mathcal{D}^2_{++0}\cdot\overline{\mathcal{D}}_{3}^{+0+}\Big|_{\tau_\phi\rightarrow\tau_\psi}
\end{pmatrix},\\
{\bf M}&=\begin{pmatrix}
1 & 0\\
\frac{(\tau_{\mathcal O}-2\tau_\psi)(\tau_{\mathcal O}+2\tau_\psi-4)}{4\ell (\tau_{\mathcal O} +\ell-1)}& \frac{(\tau_\psi-1)^{-2}}{4\ell  (\tau_{\mathcal O} +\ell-1)}
\end{pmatrix}.
\end{aligned}\end{equation}
Using~\eqref{eq:differential_operators} we can compute~\cite{Costa:2011dw} $W^a$ in~\eqref{eq:u-channel} as
\begin{equation}
W^a=\sum_{b=1}^2{\bf M}^{ab} {\bf D}^b\;W_{seed}^{(p=0)}.
\end{equation}

\subsection{Analytic Results}\label{sec: analytic bootstrap eqns}
We study the u-t and s-t bootstrap equation for the correlator~\eqref{eq:the_correlator}. We perform their analytic analysis to the leading and sub-leading orders. The sub-leading corrections will depend on the OPE coefficients of the minimal (lowest) twist operators.  Let us discuss in more details what minimal twist operator do we expect.

In the u-channel~\eqref{eq:u-channel} we can exchange only traceless symmetric bosons, thus according to~\eqref{eq: twist definition} the minimal twist operators can be light scalars, conserved $\ell=1$ currents and the energy-momentum tensor. We will denote twist and spin of these operator by $\tau_{\OO_m}$ and $\ell_{\OO_m}$.
In the s-channel~\eqref{eq:s-channel} we can exchange only fermionic operators, according to~\eqref{eq: twist definition} they could be light $(1,0)$ fermions or the super-symmetric current $(2,1)$. We will denote twist and spin of these operators by $\tau_{\Psi_m}$ and $\ell_{\Psi_m}$.

We use the ansatz~\eqref{eq:twist_behavior} and~\eqref{eq:OPE_coefficients_behavior} for twists and products of OPE coefficients respectively. We report these expressions here again for convenience
\begin{equation}\label{eq:OPE_coefficients_behavior_repeated}
\tau=\tau^0+\frac{\gamma}{\ell^D}+\ldots,\qquad
P=4^{-\ell}\;\ell^B\;\left(P^0+\frac{P^1}{\ell^C}+\ldots\right).
\end{equation}

\subsubsection{Equations: u-t-channel}\label{subsec:u-t}
We equate~\eqref{eq:u-channel} with~\eqref{eq:t-channel} and perform a change of variables according to the prescription below~\eqref{eq:omega_variables}. Schematically the bootstrap equation has the form~\eqref{eq:u-t-equation_schematic_new} and can be studied in the light-cone limit
\begin{equation}\label{eq:lict_cone_w-variable}
\omega\ll1-\overline\omega\ll1.
\end{equation}
Applying the procedure of section~\ref{sec:recipe} we have found that there are two towers of large spin fermionic operators
\begin{equation}\label{eq:towersOperators:u-t}
\Psi:\quad X^{(\ell+1,\ell)},\;\; Y^{(\ell+1,\ell)}
\end{equation}
together with their hermitian conjugates $\overline X$ and $\overline Y$.
We determine their twists together with the products of OPE coefficients
\begin{equation}\label{eq:products_OPE}
P_{X\ol X},\;P_{\ol X X},\;
P_{\ol Y Y},\;
P_{Y \ol Y}.
\end{equation}
To the leading order twists read as 
\begin{equation}\label{eq:leadingTwistXY}
\tau_{\ol X}^0= \tau_\phi+\tau_\psi+2n,\quad
\tau_{Y}^0= \tau_\phi+\tau_\psi+2n+1,
\end{equation}
where $n$ is a non-negative integer spanning the tower of operators. Both anomalous dimensions are suppressed by the spin $\ell$ with the exponent
\begin{equation}\label{eq:exponentD_ut}
D_{\ol X}= D_{Y}=\tau_{\OO_m}
\end{equation}
and have the following coefficients\footnote{In the equations below one should make a replacement $P_{T}^a\rightarrow c_{T}^{-1}\,P_{T}^a$ for the energy-momentum tensor, where $c_T>0$ is a real number which appears in the $\langle T T\rangle$ normalization. This happens because of the following. The normalization of 2-point functions is fixed only up to a positive real constant. In general we use this freedom to set the constant to 1. In the case of conserved operators we do not have such a freedom since the normalization of 2-point functions is completely fixed by the algebra they obey. To correct for this the above replacement is required.
}
\begin{align}\label{anomalous_dimension_b_1}
\gamma^b_{\ol X} &=-(i P^{a=1}_{\mathcal O_m})\;
\frac{(-1)^{\ell_{\mathcal O_m}}\;f(\tau_\phi,\tau_\psi)}{\left(\tau_\psi+1\right)_{-\frac{\tau_{\mathcal O_m}}{2}}},\\
\label{anomalous_dimension_b_2}
\gamma^b_{Y} &=-
\left(\frac{\tau_\psi-\frac{\tau_{\OO_m}}{2}-1}{\tau_\psi-1}\,(iP^{a=1}_{\OO_m})-\frac{\tau_\psi-\frac{\tau_{\OO_m}}{2}}{\tau_\psi-1}\,(iP^{a=2}_{\OO_m})\right)\;
\frac{(-1)^{\ell_{\mathcal O_m}}\;f(\tau_\phi,\tau_\psi+1)}{ \left(\tau_\psi+1\right)_{-\frac{\tau_{\mathcal O_m}}{2}}},
\end{align}
where we have defined
\begin{equation}
f(\tau_\phi,\tau_\psi)\equiv
\frac{2^{\tau_{\mathcal O_m}+2 \ell_{\mathcal O_m}} \left(\frac{\tau_{\mathcal O_m}}{2}+\ell_{\mathcal O_m}\right)_{\frac12}}{\sqrt{\pi }\;(\tau_\phi )_{-\frac{\tau_{\mathcal O_m}}{2}}}\sum_{j=0}^n\frac{(n)_j \left(\tau_\phi+\tau_\psi+n-3\right)_j  \left(\frac{\tau_{\OO_m}}{2}+\ell_{\OO_m}-j\right)_{2j}}{j!(\tau_\phi-1)_j\left(\tau_\psi-1\right)_j}.
\end{equation}
We keep $a=1$ and $a=2$ to indicate explicitly that we deal with two OPE coefficients (not to be confused with orders in our large spin perturbation theory). The subscript $b$ in the anomalous dimensions stays to indicate that it reproduces bosonic minimal-twist operators.
The products of OPE coefficients have the following $B$ exponents
\begin{equation}\label{eq:exponentsBut}
B_{\ol XX}=\tau_\phi+\tau_\psi-\frac{1}{2},\quad
B_{Y\ol Y}=\tau_\phi+\tau_\psi-\frac{3}{2}.
\end{equation}
The exponents $B_{X\ol X}$ and $B_{\ol Y Y}$ are sub-leading to~\eqref{eq:exponentsBut} and thus the products of OPE coefficients governed by this exponents are zero to the order we work in. In the leading order
\begin{equation}\label{eq:P0ut}
P^0_{\ol X X}=
-\frac{2^{-\tau^0_{\ol X}}\;g(\tau_\phi,\tau_\psi)}{\left( \tau_\phi+\tau_\psi+n-3\right)_n},\quad
P^0_{Y\ol Y}=
\frac{
	2^{-\tau^0_{Y}}\;g(\tau_\phi,\tau_\psi)(\tau_\psi-1+n)}{\left( \tau_\phi+\tau_\psi+n-2\right)_n},
\end{equation}
where we have defined
\begin{equation}
g(\tau_\phi,\tau_\psi)\equiv\frac{
	2\sqrt{\pi}\;(\tau_\phi-1)_n\left(\tau_\psi-1\right)_n}{ n!\;\gammafn{\tau_\phi}\gammafn{\tau_\psi+1}}.
\end{equation}
Finally the sub-leading order OPE coefficients are suppressed by large spin $\ell$
with the exponents
\begin{equation}\label{eq:exponentC_ut}
C_{\ol X X}= C_{Y\ol Y}=\tau_{\OO_m}
\end{equation}
and the coefficients for $n=0$ read as\footnote{We have also found expression for $n\neq 0$, however their form is not presentable in the paper. We refer the reader to the accompanying code for details.}
\begin{equation}\label{eq:OPE_corrections_b}
n=0:\qquad
P^{1\;b}_{\ol X X}=\gamma^b_{\ol X}\;h,\qquad
P^{1\;b}_{Y\ol Y}=\gamma^b_{Y}\;h,\qquad
h\equiv \psi\left( \tfrac{\tau_{\OO_m}}{2}+\ell_{\OO_m}\right)-\psi(1) -\ln2.
\end{equation}

We would like to note, that since our blocks are normalized with the factor $(-1)^\ell$ according to~\eqref{eq:Coefficients}, it was crucial to compensate this factor by another $(-1)^\ell$ coming from the analytic continuation~\eqref{eq:minusSignAnalyticContinuation} to reproduce the identity operator in our bootstrap equations.
Alternatively we could have worked with the s-t equations of another correlator
\begin{equation}\label{eq:alternative correlator}
\langle \phi(\point_1)\,\phi(\point_2)\,\psi(\point_3)\,\overline\psi(\point_4) \rangle
\end{equation}
as discussed around~\eqref{eq:alternative_approach_u-t_equations}. In this case no analytic continuation is needed, however the products of OPE coefficients $P$ will neatly organize themselves into positive (negative) definite quantities like in~\eqref{eq:positivity_of_OPE_coefficients} times a factor $(-1)^\ell$. Which means that the final equations will never have a $(-1)^\ell$ factor. This will allow to reproduce the identity operator correctly.

\subsubsection{Equations: s-t-channel}\label{subsec:s-t}
The s-t bootstrap equation is formed by equating the expression~\eqref{eq:s-channel} with~\eqref{eq:t-channel}. Schematically it has the form~\eqref{eq:typical_bootstrap_equation} and can be studied in the light-cone limit~\eqref{eq:light-cone_limit}.

Applying the procedure of section~\ref{sec:recipe}, we have found that there are again two towers of large spin fermionic operators
\begin{equation}\label{eq:towersOperators:s-t}
\Psi:\quad X'^{(\ell+1,\ell)},\;\; Y'^{(\ell+1,\ell)}
\end{equation}
together with their hermitian conjugates $\overline X'$ and $\overline Y'$. The operators~\eqref{eq:towersOperators:s-t} are in principle different from~\eqref{eq:towersOperators:u-t}, we put a prime to indicate that. As before we determine their twists together with products of OPE coefficients
\begin{equation}\label{eq:products_OPEprime}
P_{X'\ol X'},\;P_{\ol X' X'},\;
P_{\ol Y' Y'},\;
P_{Y' \ol Y'}.
\end{equation}
In the leading order the twists read as
\begin{equation}\label{eq:leadingTwistX'Y'}
\tau_{\ol X'}^0= \tau_\phi+\tau_\psi+2n,\quad
\tau_{Y'}^0= \tau_\phi+\tau_\psi+2n+1,
\end{equation}
where $n$ is a non-negative integer spanning the tower of operators.

The coefficients $P^0_{\ol X' X'}$ and $P^0_{Y' \ol Y'}$ cannot contain a $(-1)^\ell$ factor in order to reproduce correctly the absence of identity operator. We cannot however put further constraints on them, they are left undetermined. As a consequence the exponents $B_{\ol X' X'}$ and $B_{Y'\ol Y'}$ are also left undetermined.

The anomalous dimensions are suppressed by large $\ell$ with exponents $D$.  Since the exponents $B$ are unknown we can only compute the differences
\begin{equation}\label{eq:exponentD_st}
B_{\ol X'X'}-D_{\ol X'}=
\left(\tau_{\phi}+\tau_{\psi}-\frac{1}{2}\right)-\tau_{\Psi_m},\quad
B_{Y'\ol Y'}-D_{Y'}=\left(\tau_{\phi}+\tau_{\psi}-\frac{3}{2}\right)-\tau_{\Psi_m}.
\end{equation}
The coefficients of anomalous dimensions are given in terms of unknown $P^0$ as\footnote{We have computed also the values for $n\neq0$, but we decided not to present them in the paper.}
\begin{align}\label{anomalous_dimension_f_1}
n=0:\qquad\gamma^f_{\ol X'} &=(-1)^{\ell+1}
\times\frac{P_{\oPsi_m\Psi_m}}{P_{\ol X^\prime X^\prime}^0}\times
\frac{\tilde f(\tau_\phi,\tau_\psi)}{\Gamma\left(\frac{1}{2}\,(\tau_\phi+\tau_\psi-\tau_{\Psi_m}+1)\right)^2},\\
\label{anomalous_dimension_f_2}
n=0:\qquad\gamma^f_{Y'} &=(-1)^{\ell+1}\;
\times\frac{P_{\Psi_m\oPsi_m}}{P_{Y^\prime \ol Y^\prime}^0}\times
\frac{\tilde f(\tau_\phi,\tau_\psi+1)}{\;\Gamma\big(\frac{1}{2}\,(\tau_\phi+\tau_\psi-\tau_{\Psi_m})\big)^2} ,
\end{align}
where we have defined
\begin{equation}\label{eq:ftilda}
\tilde f(\tau_\phi,\tau_\psi)\equiv
\frac{2^{2-\tau_\phi-\tau_\psi}\sqrt{\pi}(-1)^{\ell_{\Psi_m}}\;\Gamma(1+2\ell_{\Psi_m}+\tau_{\Psi_m})}{
\Gamma\big(\frac{1}{2}\,(\tau_\phi-\tau_\psi+\tau_{\Psi_m}+2\ell_{\Psi_m}+1)\big)
\Gamma\big(\frac{1}{2}\,(-\tau_\phi+\tau_\psi+\tau_{\Psi_m}+2\ell_{\Psi_m}+1)\big)
}.
\end{equation}

The sub-leading corrections to products of OPE coefficients are suppressed by the large spin $\ell$ with exponents $C$. Again, we can only determine the differences
\begin{equation}\label{eq:exponentsBst}
B_{\ol X'X'}-C_{\ol X'X'}=
\left(\tau_{\phi}+\tau_{\psi}-\frac{1}{2}\right)-\tau_{\Psi_m},\quad
B_{Y'\ol Y'}-C_{Y'\ol Y'}=\left(\tau_{\phi}+\tau_{\psi}-\frac{3}{2}\right)-\tau_{\Psi_m}.
\end{equation}
The coefficients $P^1$ are then given by
\begin{equation}
\label{eq:OPE_corrections_f}
n=0:\qquad
P^{1\;f}_{\ol X' X'}=\frac{1}{2}\,\gamma^f_{\ol X'}\,P^0_{\ol X' X'} \,\tilde h(\tau_\phi,\tau_\psi),\quad
P^{1\;f}_{Y'\ol Y'}=\frac{1}{2}\,\gamma^f_{Y'\ol Y'}\,P^0_{Y'\ol Y'}\,\tilde h(\tau_\phi,\tau_\psi+1),
\end{equation}
where we have defined
\begin{equation}
\tilde h(\tau_\phi,\tau_\psi)\equiv
-2\psi(1)-\ln4
+\psi\Brk{\frac{\tau_\phi-\tau_\psi+\tau_{\Psi_m}+1}{2}+\ell_{\Psi_m}}
+\psi\Brk{\frac{-\tau_\phi+\tau_\psi+\tau_{\Psi_m}+1}{2}+\ell_{\Psi_m}}.
\end{equation}

\subsubsection{Summary}\label{sec:summary}
In previous sections we found that to satisfy the bootstrap equations in the u-t channel and in the s-t channel we are required to have operators $(X,Y)$ and $(X',Y')$ respectively. They look very similar: their leading twists~\eqref{eq:leadingTwistXY} and~\eqref{eq:leadingTwistX'Y'} are identical and the structure of exponents~\eqref{eq:exponentD_ut}, \eqref{eq:exponentsBut}, \eqref{eq:exponentC_ut} and \eqref{eq:exponentD_st}, \eqref{eq:exponentsBst} is strikingly similar. It is thus tempting to identify the pair $(X,Y)$ with $(X',Y')$.

Under the assumption that the identity operator in the u-t channel is fully reproduced only by the operators $(X,Y)$ we can argue as follows. Since the operators $(X',Y')$ cannot contribute to the u-channel one is forced to set $P_{\ol X^\prime X^\prime}^0=P_{Y^\prime \ol Y^\prime}^0=0$, because these coefficients do not have a $(-1)^\ell$ factor as was argued below~\eqref{eq:ftilda} and would give a non-zero contribution otherwise.
However this requirement is inconsistent with the coefficients of anomalous dimensions~\eqref{anomalous_dimension_f_1}, \eqref{anomalous_dimension_f_2} which would immediately blow up. Thus we conclude that $(X,Y)=(X',Y')$.

We can then neatly combine the results of sections~\ref{subsec:u-t} and~\ref{subsec:s-t} as follows. The twists of large spin fermionic operators read as
\begin{align}
\label{eq:final_summary_1}
&\tau_{X}= \tau_{\overline X}=\left( \tau_\phi+\tau_\psi+2n\right)
+\frac{\gamma_{\ol X}^{b}}{\ell^{\tau_{\OO_m}}}
+\frac{\gamma_{\ol X}^{f}}{\ell^{\tau_{\Psi_m}}}, \\
\label{eq:final_summary_2}
&\tau_{Y}=\tau_{\overline Y}= \left(\tau_\phi+\tau_\psi+2n+1\right)
+\frac{\gamma_{Y}^{b}}{\ell^{\tau_{\OO_m}}}
+\frac{\gamma_{Y}^{f}}{\ell^{\tau_{\Psi_m}}},
\end{align}
where the coefficients of anomalous dimensions $\gamma^b$ are given in~\eqref{anomalous_dimension_b_1}, \eqref{anomalous_dimension_b_2} and $\gamma^f$ are given in~\eqref{anomalous_dimension_f_1}, \eqref{anomalous_dimension_f_2}.
The OPE coefficients have the following form
\begin{align}\label{eq:final_summary_3}
P_{\ol X X} &=
4^{-\ell}\ell^{\tau_\phi+\tau_\psi-\frac{1}{2}}
\left(P^0_{\ol X X}
+\frac{P^{1\,b}_{\ol X X}}{\ell^{\tau_{\OO_m}}}
+\frac{P^{1\,f}_{\ol X X}}{\ell^{\tau_{\Psi_m}}}\right),\\
\label{eq:final_summary_4}
P_{Y\ol Y} &=4
^{-\ell}\ell^{\tau_\phi+\tau_\psi-\frac{3}{2}}
\left(P^0_{Y\ol Y}
+\frac{P^{1\,b}_{Y\ol Y}}{\ell^{\tau_{\OO_m}}}
+\frac{P^{1\,b}_{Y\ol Y}}{\ell^{\tau_{\Psi_m}}}\right),
\end{align}
where $P^0_{\ol X X}$ and $P^0_{Y\ol Y}$ are given in~\eqref{eq:P0ut}, $P^{1\,b}_{\ol X X}$ and $P^{1\,b}_{Y\ol Y}$ are given in~\eqref{eq:OPE_corrections_b}, $P^{1\,f}_{\ol X X}$ and $P^{1\,f}_{Y\ol Y}$ are given in~\eqref{eq:OPE_corrections_f} and
\begin{equation}
P_{X\ol X} = P_{\ol Y Y} = 0
\end{equation}
to the same order of expansion.

Looking at the final result~\eqref{eq:final_summary_1}-\eqref{eq:final_summary_4} one notices that all the ``fermionic'' quantities (labeled by a superscript $f$) contain a factor $(-1)^\ell$ and thus their contribution to the u-t bootstrap equation completely vanishes, while all the ``bosonic'' quantities (labeled by a superscript $b$) do not have a $(-1)^\ell$ factor and thus their contribution completely vanishes in the s-t equation.

\subsection{Generalized Free Theory}\label{sec:GFT}
Here we consider a GFT with two primary fundamental fields: a real scalar $\phi$ and a fermion $\psi$. See footnote~\ref{foot:definition_GFT} for the definition of GFTs.
All the double-twist composite primary operators $\Psi^{(\ell,\bar{\ell})}$~\footnote{We have suspended here the convention~\eqref{eq:convention} for $\Psi^{(\ell,\bar{\ell})}$ for simplicity of narration.} are constructed from $\phi$, $\psi$ and their derivatives. We will not need their precise expressions and instead are interested only in their schematic form. For that consider a 3-point function
\begin{equation}\label{eq:3_point_function_GFT}
\langle\phi(\point_1)\psi(\point_2)\Psi^{(\ell,\bar{\ell})}(\point_3).
\end{equation}
It can be computed using Wick contractions and give a non vanishing result only if the operators $\Psi^{(\ell,\bar{\ell})}$ are of the following two types
\begin{align}\label{eq:schematic_fermionic_oeprators_1}
\overline{\mathcal{X}}^{(\ell,\,\ell+1)}(\point_3) &=
\Big(s_3\sigma \partial\bar s_3\Big)^\ell[\partial^2]^n\,\phi(x_3)\Big(\bar s_3^{\dot\alpha}\bar\psi_{\dot\alpha}(x_3)\Big),\\
\label{eq:schematic_fermionic_oeprators_2}
\mathcal{Y}^{(\ell+1,\,\ell)}(\point_3) &=
\Big(s_3\sigma \partial\bar s_3\Big)^\ell[\partial^2]^n\,\phi(x_3)\Big( s_3^{\alpha}\sigma_{\alpha\dot\beta}^\mu\partial_\mu\bar\psi^{\dot\beta}(x_3)\Big).
\end{align}
We choose the curly font for the operators $(\mathcal{X},\mathcal{Y})$ in order to distinguish them from the operators~\eqref{eq:towersOperators:u-t} and~\eqref{eq:towersOperators:s-t}.
The twists of these operators are easy to read off from their schematic expressions
\begin{equation}\label{eq:GFT_scaling_dimensions}
\tau_{\overline{\mathcal{X}}} =\tau_\phi+\tau_\psi+2n,\quad
\tau_{\mathcal{Y}} =\tau_\phi+\tau_\psi+2n+1.
\end{equation}
In GFT one can also compute the 4-point function~\eqref{eq:TheCorrelationFunction} using Wick contractions
\be\label{wickcont}
\langle \phi(\point_1)\psi(\point_2)\phi(\point_3)\bar\psi(\point_4) \rangle =
\langle \phi(\point_1)\phi(\point_3)\rangle \; \langle\psi(\point_2)\bar\psi(\point_4) \rangle=
\frac{- i\;\hat{ \mathbb{I}}^{42}}{(x_{13}^2)^{\tau_\phi}(x_{24}^2)^{\tau_\psi+1}}.
\ee
It is clear that in OPE of the 4-point function~\eqref{eq:TheCorrelationFunction} only the double-twist operators~\eqref{eq:schematic_fermionic_oeprators_1} and~\eqref{eq:schematic_fermionic_oeprators_2} can be exchanged. Thus, given the explicit expression for the 4-point functions~\eqref{wickcont}, the explicit values of twists of exchanged operators~\eqref{eq:GFT_scaling_dimensions}, one can use the s-, t- and u- channel expansions~\eqref{eq:s-channel},~\eqref{eq:t-channel} and~\eqref{eq:u-channel} in conformal blocks to compute products of OPE coefficients in the GFT. 

\paragraph{u-channel} is the simplest case since the identity operator (the first term in~\eqref{eq:u-channel}) reproduces~\eqref{wickcont} precisely. Thus there are no other bosonic operators in our GFT which can be exchanged in the 4-point function~\eqref{eq:TheCorrelationFunction}.

\paragraph{s-channel} suites the best for determining the OPE coefficients of the operators~\eqref{eq:schematic_fermionic_oeprators_1} and~\eqref{eq:schematic_fermionic_oeprators_2}.
Equating the s-channel expansion~\eqref{eq:s-channel} with~\eqref{wickcont} and splitting the obtained expression into independent tensor structures we get two equations of the schematic form
\begin{equation}\label{eqn: GFT s-ch expan}
\left(z\bar{z}\right)^{\#}\delta^I_1=\sum_{\ell=0}^{\infty}\sum_{n=0}^{\infty}f^I_{\ell,n}(z,\bar z),
\end{equation}
where $\delta$ is the Kronecker symbol and $f^I_{\ell,n}(z,\bar z)$ represents conformal blocks together with the OPE coefficients.
Expanding these equations around $z=0$ and $\bar z=0$ and equating coefficients in front of the same powers of $z$ and $\bar z$ on both sides, one gets a system of linear equations for the coefficients $P_{\overline{\mathcal{X}}\mathcal{X}}$ and $P_{\mathcal{Y}\ol{\mathcal{Y}}}$. Only a finite number of coefficients $P$ contribute at each order of expansion, thus, the system is exactly solvable. We found a solution for several values of $\ell$ and $n$ and guessed the generic solution, which is
\begin{equation}
\begin{aligned}\label{eq:OPE_GFT}
P_{\overline{\mathcal{X}}\mathcal{X}}^{GFT}(n,\ell) =
&-\frac{(\tau_\phi-1)_n\,(\tau_\psi-1)_n}{n!\,(\tau_\phi+\tau_\psi+n-3)_n}\times\\
&\frac{(\tau_\phi)_{\ell+n}\,(\tau_\psi+1)_{\ell+n}}
{\ell!\,(\ell+2)_n\,(\tau_\phi+\tau_\psi+\ell+n	-1)_n\,(\tau_\phi+\tau_\psi+\ell+2n)_\ell},\\	
P_{\mathcal{Y}\overline{\mathcal{Y}}}^{GFT}(n,\ell) =		
&+\frac{(\tau_\phi-1)_n\,(\tau_\psi-1)_{n+1}}
{n!\,(\tau_\phi+\tau_\psi+n-2)_n}\times\\
&\frac{(\tau_\phi)_{\ell+n+1}\,(\tau_\psi+1)_{\ell+n}}
{\ell!\,(l+2)_{n+1}(\tau_\phi+\tau_\psi+\ell+n-1)_{1+n}\,
	(\tau_\phi+\tau_\psi+\ell+2n+1)_\ell}
\end{aligned}
\end{equation}
together with
\begin{equation}
P_{\mathcal{X}\overline{\mathcal{X}}}^{GFT}(n,\ell) = 
P_{\overline{\mathcal{Y}}\mathcal{Y}}^{GFT}(n,\ell) = 0.
\end{equation}
Notice that signs of the above expression neatly match the signs in~\eqref{eq:positivity_of_OPE_coefficients} providing a nice consistency check.

\paragraph{t-channel} is very similar to the s-channel. We can equate the t-channel expansion~\eqref{eq:t-channel} with~\eqref{wickcont}, split the resulting expression in two independent tensor structures and get two equations of the schematic form
\begin{equation}\label{eq:t_equation}
\cancel{\left(z\bar{z}\right)^{\#}}\delta^I_1=
\sum_{\ell=0}^{\infty}\sum_{n=0}^{\infty}\cancel{\left(z\bar{z}\right)^{\#}}\Big((1-z)(1-\bar{z})\Big)^{\#}  \tilde f^I_{\ell,n}(1-z,1-\bar z)
\end{equation}
for the OPE coefficients.
These equations are apparently very different from~\eqref{eqn: GFT s-ch expan}\footnote{Modulo the obvious $z\leftrightarrow1-z$ and $\zb \leftrightarrow1-\zb$ exchange, we have a multiplier coming from the kinematic factor and we have a mixing of conformal blocks.} but are in fact equivalent.
Expanding~\eqref{eqn: GFT s-ch expan} around  $z=\bar z=1$ one finds a solution for the OPE coefficients which is identical to~\eqref{eq:OPE_GFT}. This is another nice consistency check.

\paragraph{Connection to the Analytic Bootstrap}
Taking the leading behavior of~\eqref{eq:final_summary_3}, \eqref{eq:final_summary_4} and comparing it with~\eqref{eq:OPE_GFT} one can notice that the following holds
\begin{equation}\label{eq:connection_GFT}
4^{-\ell}\ell^{\tau_\phi+\tau_\psi-\frac{1}{2}}P^0_{\ol X X}=
\lim_{\ell\rightarrow\infty}P_{\ol{\mathcal X}\mathcal X}^{\text{GFT}},\quad
4^{-\ell}\ell^{\tau_\phi+\tau_\psi-\frac{3}{2}}P^0_{Y\ol Y}=
\lim_{\ell\rightarrow\infty} P^{\text{GFT}}_{\mathcal{Y}\overline{\mathcal{Y}}}.
\end{equation}
This happens because the equations for the GFT OPE coefficients match the bootstrap equations in the light-cone limit. More precisely, take the $t$-channel GFT equation~\eqref{eq:t_equation} and change variables according to~\eqref{eq:omega_variables} and the prescription below it.
This expression will then be equivalent to~\eqref{eq:u-t-equation_schematic_new} in the light-cone limit~\eqref{eq:lict_cone_w-variable}. Thus both equations should have the same asymptotic solution.

\newpage

\section{Conclusions}\label{sec:discussion}
In this paper we discuss the analytic bootstrap approach~\cite{Fitzpatrick:2012yx,Komargodski:2012ek} to general 4-point functions in 4D CFTs~\cite{Elkhidir:2014woa,Echeverri:2015rwa,Echeverri:2016dun,Cuomo:2017wme}. To simplify computations in practice we develop and attach to the paper a little Mathematica code ``analyticBootstrap4D.m''.
We proceed by studying in detail the s-t and u-t bootstrap equations of the scalar-fermion correlator $\langle\phi\,\psi\,\phi\,\overline\psi\rangle$ and show how the results from these two channels complement each other. We find that for every pair of scalar $\phi$ and fermion $\psi$ operators in the spectrum correspond two infinite towers of large spin fermionic operators. We compute their twist and products of OPE coefficients in the leading and sub-leading order~\eqref{eq:final_summary_1}-\eqref{eq:final_summary_4}.
The sub-leading corrections are governed by the minimal-twist operators in the CFT. They can be both bosonic (light scalars, conserved currents and the energy-momentum tensor) and fermionic (light $(1,0)$ and $(0,1)$ fermions and the supersymmetric current). We provide details of all the computations in the attached Mathematica file "Example.nb".

To complement the discussion we have also studied the scalar-fermion GFT. We derive closed-form expression for the GFT data~\eqref{eq:GFT_scaling_dimensions} and~\eqref{eq:OPE_GFT}. We show that in  the large spin limit the GFT data perfectly matches the leading twist and OPE coefficients obtained using the analytic bootstrap.

In the same way of this paper one can study other spinning correlators. To complement numerical bootstrap studies in future it might be interesting to consider correlators of the form  $\langle\bar\psi\psi\bar\psi\psi\rangle$ and $\langle J J J J \rangle$, where $J$ is the conserved current. One can also study $\langle \phi \mathcal J \bar{\mathcal J} \phi\rangle$, where $\mathcal J$ is (2,1) fermion (spin 3/2 fermion). In case  $\mathcal J$ is conserved, analogously to~\cite{Li:2015itl,Hofman:2016awc} one can access the supersymmetric conformal collider bounds.

The results of section~\ref{sec:bootstrap_equations} can be interpreted in the context of AdS. One can say that the double-twist operators $X$ and $Y$ on the CFT side are dual to a system of largely separated scalar and fermion particle orbiting each other with the large relative angular momentum $\ell$.\footnote{This system will also have two states $\ell\otimes\frac{1}{2}=(\ell-\frac{1}{2})\oplus(\ell+\frac{1}{2})$. Here we use half-integer values to label representations. It is different from the rest of the paper where we use only integer values.}
The anomalous dimensions of operators $X$ and $Y$ would then correspond in AdS to energy shifts of the original system caused by interactions between orbiting particles. For our system we have two types of such interactions bosonic attractive/repulsive forces (gravity, Higgs-like force) or fermionic interactions which cause swapping of orbiting particles. It would be interesting to compute the energy shifts due to these interactions in AdS~\cite{Fitzpatrick:2010zm,Fitzpatrick:2014vua}\footnote{In doing that techniques similar to~\cite{Costa:2014kfa} might be of help.}
and compare them to anomalous dimensions found in this work.
Here we only notice that the anomalous dimensions induced by the energy-momentum tensor are negative. This can be easily seen by plugging~\eqref{eq:wardIdentitiesStressTensor} into~\eqref{anomalous_dimension_b_1} and~\eqref{anomalous_dimension_b_2}, indicating that gravity in AdS between scalar and fermion particle is attractive.

Recent discovery of the Lorentizan inversion formula~\cite{Caron-Huot:2017vep} for scalars brings analytic studies to a new level. Its generalization to spin is a matter of time. Our results might serve as a benchmark for checking future spinning inversion formula in 4D.

\section*{Acknowledgments} The authors are grateful to Marco Serone for collaboration at the initial stage of this project and for numerous discussions later on. We thank Agnese Bissi, Gabriel Cuomo, Jared Kaplan, Petr Kravchuk, Dalimil Mazac, David Meltzer, Jo\~ao Penedones, David Poland, David Simmons-Duffin and Matthew Walters for useful discussions.
DK is supported by the Simons Foundation grant 488649 (Simons collaboration on the non-perturbative bootstrap) and by the National Centre of Competence in Research SwissMAP funded by the Swiss National Science Foundation. The research of EE is supported in part by the Knut and Alice Wallenberg Foundation under grant Dnr KAW 2015.0083.

\appendix
\section{Ward Identities}\label{sec:Ward_identities}
As a consequence of conformal symmetry there exist a conserved traceless symmetric spin two operator, the energy momentum tensor $T^{\mu\nu}(x)$. In radial quantization\footnote{To work with radial quantization we need to analytically continue all our operators to Euclidean signature. As a result we get an overall factor $-1=(-i)^2$ in~\eqref{eq:conserved_charge_general}, see~\cite{Poland:2010wg}.} one can form 15 special non-local operators (conserved charges) $Q_\epsilon$~\cite{Simmons-Duffin:2016gjk,Penedones:2016voo}
\begin{equation}\label{eq:conserved_charge_general}
Q_\epsilon\equiv -\int|x|^3\,dS^3\,\hat x_\mu\;\Big(\epsilon_\nu(x)
\; T^{\mu\nu}(x)\Big),\qquad
\hat x^\mu\equiv\frac{x^\mu}{|x|},\qquad
\int dS^3=2\pi^2,
\end{equation}
where $\epsilon_\nu(x)$ is the Killing vector corresponding to the most general infinitesimal conformal transformation $x_\nu\rightarrow x_\nu+\epsilon_\nu(x)$ which has the form
\begin{equation}\label{eq:infinitesimal_conformal_transformation}
\epsilon_\nu(x)\equiv a_\nu+\lambda x_\nu +\omega_{[\nu\rho]}x^\rho+
\Big(b_\nu x^2 - 2 (b\cdot x)x_\nu\Big).
\end{equation}
Here $a_\nu,\;\lambda,\;\omega_{[\nu\rho]}$ and $b_\nu$ are some constants parameters associated to translations, dilatation, rotations and special conformal transformations respectively. We denote the associated conserved charges $Q_\epsilon$ by $P^\mu$, $D$, $M^{\mu\rho}$ and $K^\mu$. They form the conformal algebra and give rise to commutation relations with primary operators, see (A.14)-(A.17) in~\cite{Cuomo:2017wme}. By injecting the charges $Q_\epsilon$ between two states and evaluating them using~\eqref{eq:conserved_charge_general} on one side and commutation relations on  another side, one arrives at Ward identities.

\paragraph{Tensor vs. spinor notation} Conserved bosonic traceless symmetric currents are naturally defined in tensor language. To make connection to operators in spinor formalism used in this work we need to provide a map between tensor and spinor notation. This map is not uniquely defined, we make the following choice
\begin{equation}
\OO(x)^{\alpha_1\ldots\alpha_\ell;\;}{}^{\dot\beta_1\ldots\dot\beta_{\ell}}=
\bar\sigma_{\mu_1}^{(\dot{\beta}_1\alpha_1}\ldots\bar\sigma_{\mu_\ell}^{\dot{\beta}_\ell\alpha_\ell)}\times
\OO(x)^{\mu_1\ldots\mu_\ell},
\end{equation}
where symmetrization is performed for both sets of dotted and undotted indices independently.
It has the following inverse operation
\begin{equation}\label{eq:spinor_to_tensor}
\OO(x)^{\mu_1\ldots\mu_\ell}\equiv\frac{(-1)^\ell}{2^\ell}\times
\sigma^{\mu_1}_{\alpha_1\dot{\beta}_1}\ldots\sigma^{\mu_\ell}_{\alpha_\ell\dot{\beta}_\ell}\times
\OO(x)^{\alpha_1\ldots\alpha_\ell;\;}{}^{\dot\beta_1\ldots\dot\beta_{\ell}}.
\end{equation}
Combined with an index-free spinor notation
\begin{equation}\label{eq:index_full}
\OO(x)^{\alpha_1\ldots\alpha_\ell;\;}{}^{\dot\beta_1\ldots\dot\beta_{\ell}}=
\frac{(-1)^\ell}{\ell!^2}\,\prod_{i=1}^\ell\prod_{j=1}^{\ell}
[\partial_{s}]^{\alpha_i}
[\partial_{\bar s}]^{\dot\alpha_j}\;
\OO(x,s,\bar s),
\end{equation}
defined in equation~(2.3) in~\cite{Cuomo:2017wme}, we arrive at a compact formula\footnote{We have defined the following short-hand notation for derivatives in auxiliary spinors
\begin{equation}
[\partial_{s}]^\alpha\equiv\frac{\partial}{\partial s_\alpha},\quad
[\partial_{\bar s}]^{\dot\alpha}\equiv\frac{\partial}{\partial s_{\dot\alpha}},\quad
[\partial_{s}]_\alpha\equiv-\frac{\partial}{\partial s^\alpha},\quad
[\partial_{\bar s}]_{\dot\alpha}\equiv-\frac{\partial}{\partial s^{\dot\alpha}}.
\end{equation}
The extra minus signs in the above expressions are needed to define the standard raising and lowering procedure in a consistent way. We also define coordinates in spinor notations as
$x_{{\alpha}\dot\beta}\equiv x_\mu\sigma^\mu_{{\alpha}\dot\beta}$ and $\bar x^{\dot{\alpha}\beta}\equiv x^\mu\bar\sigma_\mu^{\dot{\alpha}\beta}$. 
}
\begin{equation}\label{eq:spinor_to_vector_final}
\OO(x)^{\mu_1\ldots\mu_\ell}\equiv\frac{1}{2^\ell\;\ell!^2}\;
\prod_{i=1}^\ell(\partial_s\sigma^{\mu_i}\partial_{\bar s})\OO(x,s,\bar s).
\end{equation}
Using~\eqref{eq:spinor_to_vector_final} and the definitions of conformal conserved charges~\eqref{eq:conserved_charge_general} together with~\eqref{eq:infinitesimal_conformal_transformation} one can write for example\footnote{Analogous expression can also be written for $M^{\mu\nu}$ and $K^\mu$.}
\begin{align}
\label{eq:D}
D              &= -\frac{1}{16}
\int |x|^2\,dS^3\;(\partial_s x\partial_{\bar s})^2\;T(x,s,\bar s),\\
\label{eq:P}
P^\mu          &=-\frac{1}{16}
\int |x|^2\,dS^3\;(\partial_s \sigma^\mu\partial_{\bar s})(\partial_s x\partial_{\bar s})\;T(x,s,\bar s).
\end{align}

\paragraph{Applications}
Consider first a  scalar correlator with a traceless symmetric operator~\eqref{eq:def_5}. It can be used to derive an OPE which in the leading order reads as
\begin{equation}\label{eq:ope_scalars}
\OO^{(\ell,\ell)}_\Delta{(x,s,\bar s)}\;\phi(0)=\\
(-1)^\ell\lambda_{\langle \OO\phi\bar \phi\rangle}\;\frac{(\bar s \bar x s)^{\ell}}{x^{\Delta+l}}\;\phi(0)+\ldots.
\end{equation}
In case $\OO^{(\ell,\ell)}_\Delta$ is the energy-momentum tensor, the Ward identities for the charge $D$ lead to 
\begin{eqnarray}
\lambda_{\langle T\phi\phi\rangle}=\lambda_{\langle \phi\phi T\rangle}=-\frac{2\Delta_\phi}{3\pi^2},
\end{eqnarray}
where we have used the commutation relation $[D,\phi(0)]=\Delta_\phi\;\phi(0)$. The OPE in the case of fermion correlator~\eqref{eq:def_6} reads as
\begin{multline}\label{eq:ope_fermions}
\OO^{(\ell,\ell)}_\Delta{(x,s,\bar s)}\;\psi(0,t)=\\
i\,(-1)^\ell\,\frac{(\bar s \bar x s)^{\ell-1}}{x^{\Delta+l}}\times
\bigg(\lambda_{\langle\OO\psi\bar\psi\rangle} ^1(\bar s \bar x s)
+\lambda_{\langle\OO\psi\bar\psi\rangle}^2(\bar s \bar x t)(s\cdot\partial_{t})   \bigg)
\;\psi(0,t)+\ldots.
\end{multline}
Analogously to the scalar case, if $\OO^{(\ell,\ell)}_\Delta$ is the energy-momentum tensor, the charge $D$ gives one constraint on the OPE coefficients
\begin{equation}
2\lambda_{\langle\OO\psi\bar\psi\rangle} ^1-\lambda_{\langle\OO\psi\bar\psi\rangle} ^2=\frac{4i\Delta_\psi}{3\pi^2},
\end{equation}
where we have used $[D,\psi(0,t)]=\Delta_\psi\;\psi(0,t)$. One can verify that the result~\eqref{eq:wardIdentitiesStressTensor} obeys this constraint.
To completely fix both OPE coefficients the charge $D$ is obviously not enough and one should use other conserved charges, for example $P_\mu$. Unfortunately its action on the leading OPE term~\eqref{eq:ope_fermions} vanishes and one should extend the OPE to sub-leading order. We find however that working with full correlation function at this point is more straightforward.
We do not report the details, but only the final result~\eqref{eq:wardIdentitiesStressTensor}.

\bibliographystyle{JHEP}
\bibliography{references}

\end{document}